\documentclass[aip,rsi,amsmath,reprint,amssymb,groupedaddress]{revtex4-1}

\usepackage{graphicx}
\usepackage{siunitx}
\usepackage{amssymb}

\begin{document}

\title{Microwave probing of bulk dielectrics using superconducting coplanar resonators in distant-flip-chip geometry}

\author{Lars~Wendel}
\author{Vincent~T.~Engl}
\author{Gabriele~Untereiner}
\author{Nikolaj~G.~Ebensperger}
\author{Martin~Dressel}
\author{Ahmed~Farag}
\author{Monika~Ubl}
\author{Harald~Giessen}
\author{Marc~Scheffler}
\email[]{scheffl@pi1.physik.uni-stuttgart.de}
\affiliation{Physikalisches Institut, Universit\"at Stuttgart, 70569 Stuttgart, Germany}

\date{\today}

\begin{abstract}
Dielectric measurements on insulating materials at cryogenic temperatures can be challenging, depending on the frequency and temperature ranges of interest.
We present a technique to study the dielectric properties of bulk dielectrics at GHz frequencies. A superconducting coplanar Nb resonator is deposited directly on the material of interest, and this resonator is then probed in distant-flip-chip geometry with a microwave feedline on a separate chip. Evaluating several harmonics of the resonator gives access to various probing frequencies, in the present studies up to 20~GHz. We demonstrate the technique on three different materials (MgO, LaAlO$_3$, and TiO$_2$), at temperatures between 1.4~K and 7~K.
\end{abstract}

\maketitle

\section{Introduction}
The dielectric properties of various insulating materials are highly interesting and topic of ongoing research.\cite{Dimos1998,Brand2002,Hess1982,Mueller1979,Hering2007,Gorshunov2016} However, there is no single generic method to probe dielectrics at certain frequencies or low temperatures, and different high frequency techniques can access the desired properties in the GHz range.\cite{Krupka2006} Especially, resonant techniques are suited, since they provide good accuracy.\cite{Jacob2002,Kent1994,Krupka1994} The goal of this study is to characterize bulk dielectrics with a new resonant approach by probing multiple discrete GHz frequencies at cryogenic temperatures.

Planar superconducting microwave resonators play an important role in solid state research, including cryogenic spectroscopic studies.\cite{Frunzio2005,Hammer2007,Zmuidzinas2012,Ebensperger2016,Scheffler2015,Scheffler2013,Oliver2013,Gruenhaupt2018,Kroll2019} 
One can simultaneously probe various harmonics of a fundamental resonance frequency and perform measurements at low temperatures, even in setups with limited space for the sample probe, such as in high magnetic fields or at mK temperatures.\cite{Kubo2010,Driessen2012,Scheffler2013,Ghirri2015,Thiemann2018a,LevyBertrand2019}
Superconducting resonators can reach very high quality factors \cite{Megrant2012} and thus high sensitivity for the dielectric under study.\cite{OConnell2008,Ebensperger2019,Wisbey2019}
A further advantage is the well-established lithographic fabrication which allows straightforward modification of the resonator design to adjust to specific frequency or geometry demands. 
The approach presented here also eliminates a common problem of related experiments, namely air gaps, which strongly reduce accuracy or even inhibit completely the desired measurements.\cite{BakerJarvis2010,Krupka1999,BakerJarvis2006,Ebensperger2019}
In our approach, the superconducting coplanar waveguide (CPW) resonator is directly deposited on top of the bulk sample under study and then excited with an external microwave feedline. We chose the superconductor Nb as resonator material since its fabrication is well established,\cite{Farag2019} such resonators have already been studied in great detail \cite{Bothner2012a,Goetz2016}, and their properties can be optimized depending on the performance parameters of interest.\cite{Goeppl2008,Bothner2012b,Rausch2018}

\begin{figure}
	\centering
	\includegraphics[width=\columnwidth]{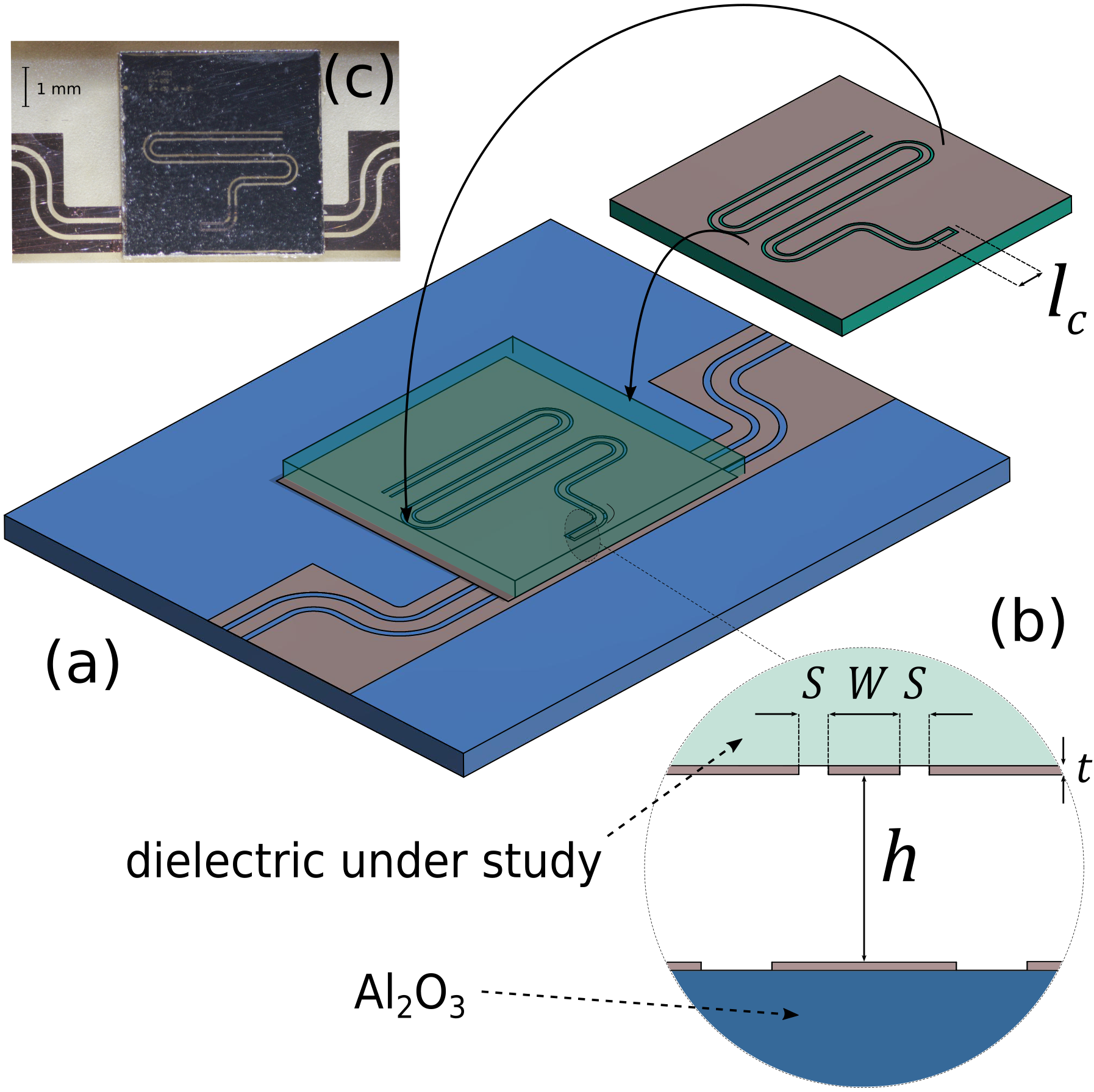}
	\caption{(a) Schematic of distant-flip-chip geometry. The sample (transparent green) with deposited CPW $\lambda / 4$-resonator is flipped over a coplanar transmission line on a substrate (blue) but kept at a distance $h$. (b) Schematic cross section of CPW, indicating distance $h$ between chip and substrate, inner conductor width $W$, gap $S$ between inner and outer conductors, and thickness $t$ of the resonator; $l_c$ is the coupling length. (c) Photograph of distant-flip-chip arrangement.}
	\label{Fig:Distant_flip-chip}
\end{figure}

\section{Resonator design: Distant-flip-chip geometry}
In Fig.\ \ref{Fig:Distant_flip-chip} the distant-flip-chip geometry is displayed.
The superconducting Nb coplanar $\lambda / 4$-resonator is fabricated with thickness $t$, center conductor width $W$, and gap width $S$ on the studied sample. 
The sample is then flipped and placed over a coplanar feedline, made of Cu on a sapphire substrate, in a certain distance $h$, such that both conductive layers are facing each other. 
The outer conductor of the feedline is designed such that most of the feedline chip surface facing the resonator chip is not covered by the Nb.
The resonator has a coupling arm with length $l_c$ and with a closed end, which is centrally placed above the center conductor of the feedline, while the opposite resonator end is open. As displayed in Fig.\ \ref{Fig:Distant_flip-chip} the resonator is shaped with a meander structure which increases its total length and therefore reaches a smaller fundamental resonance frequency. Via inductive coupling the microwave signal is transmitted from the feedline, and the $\lambda / 4$-resonator is excited. 
The distant-flip-chip design has the advantage that the microwave signal couples through the open space between sample and feedline chips, thus preventing losses and depolarization effects. The distance $h$ and the coupling length $l_c$ are of crucial importance regarding the microwave coupling and thus functionality and performance of the distant-flip-chip geometry.

There are some challenges with this approach, especially concerning control of the coupling into the resonator, certain sample requirements, and the mounting of the resonator chip. The signal may not couple sufficiently into the resonator to excite the fundamental and higher harmonic modes since the permittivity of the sample is unknown, and therefore the impedance of the resonator cannot be adjusted correctly. Furthermore, the mounting of the chip influences the coupling into the resonator.  
The distance must be set appropriately to optimize the coupling into the resonator. Additionally, the sample needs to be flat with a polished surface and size of at least a few mm to apply the distant-flip-chip geometry.

\begin{figure}
	\centering
	\includegraphics[width=\columnwidth]{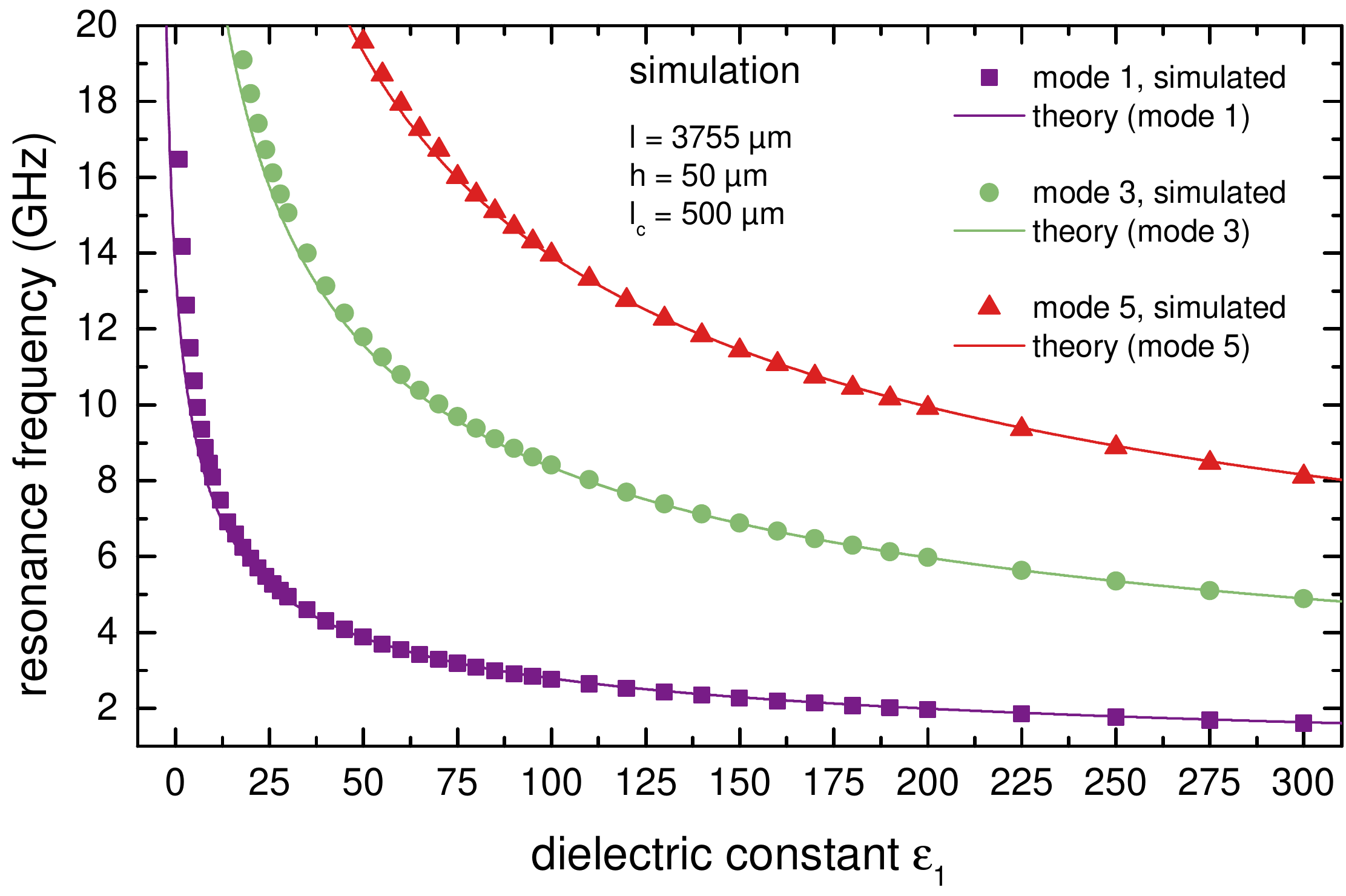}
	\caption{Fundamental and harmonic resonance frequencies of a distant-flip-chip geometry for sample permittivities up to 300. The distance of the chip is  $h = \SI{50}{\micro m}$ and the coupling length of the resonator is $l_c =~\SI{500}{\micro m}$. Symbols indicate simulation results whereas full lines are obtained analytically from conformal mapping theory.}
	\label{Fig:Sim_vs_Theo}
\end{figure}

\section{Simulations and Theory}
The distant-flip-chip geometry is checked with simulations for its suitability using \textit{CST Microwave Studio}, and the results can be compared to theoretical predictions obtained with the conformal mapping theory. \cite{Simons2001,Ghione1987} From the latter, closed-form expressions for the effective dielectric constant $\epsilon_\text{eff}$ are obtained. Within this technique the CPW is split into partial regions and the electric field is assumed to fill each of them homogeneously. Then the capacitance of each part is found and the sum gives the total capacitance. The  effective dielectric constant $\epsilon_\text{eff}$ is then calculated by the ratio of the total capacitance of the CPW and the capacitance in absence of all dielectrics. Assumptions for this technique are that the conductors have perfect conductivity, the structure is lossless, the dielectrics are isotropic, and the electric field fills each partial region perfectly. 

In Fig.\ \ref{Fig:Sim_vs_Theo} the resonance frequency as a function of the sample permittivity is displayed, obtained either by simulations or analytical calculation.
For the latter, the resonance frequencies $\nu_n$ decrease with increasing permittivity following
\begin{align}
\nu_n = \frac{nc}{4l\sqrt{\epsilon_\text{eff}}} \ , \label{Eq:ResFreq}
\end{align}
with $n$ being the number of mode, $c$ the vacuum speed of light, $l$ the length of the resonator and $\epsilon_\text{eff}$ the effective dielectric constant of the resonant design. Since the distant-flip-chip geometry utilizes a $\lambda / 4$-resonator, only odd multiples ($n = 3,5, ...$) of the fundamental frequency ($n=1$) can be excited. The simulated results, marked in Fig.\ \ref{Fig:Sim_vs_Theo} as symbols, match the analytical predictions marked as full lines. The permittivity of the sample dominates $\epsilon_\text{eff}$, which results in a shift to smaller resonance frequencies for an increasing permittivity. 

\begin{figure}
	\centering
	\includegraphics[width=\columnwidth]{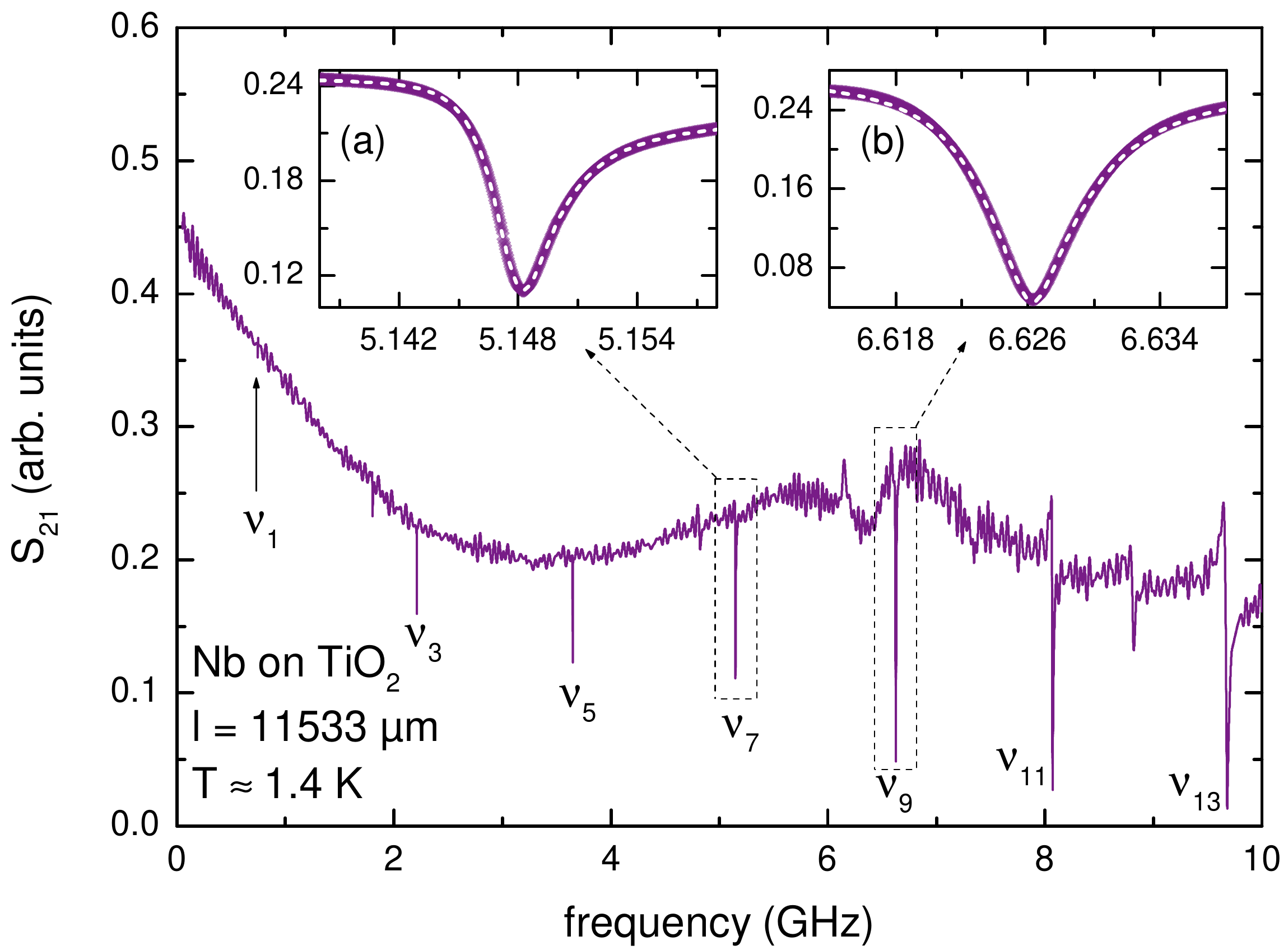}
	\caption{Transmission spectrum of a resonator chip with $l =~\SI{11533}{\micro m}$ deposited on a TiO$_2$ sample up to 10 GHz. The insets (a), (b) show two respective dips with higher resolution and the according Lorentzian fits.}
	\label{Fig:Spectra}
\end{figure}

\section{Measurements and Results}
Three different dielectric materials are used to test the approach: MgO, LaAlO$_3$, and TiO$_2$. The dielectric response of each of them at cryogenic temperatures and GHz frequencies is basically constant as a function of temperature and frequency,\cite{Krupka1994_2,Konaka1991,Sabisky1962,Parker1961} and with $\epsilon$ values roughly around 10, 25, and 200, they cover the range that is typically encountered for crystalline solids at low temperatures.
Coplanar Nb resonators are deposited onto the three substrate materials, and they are then mounted with vacuum grease in distant-flip-chip geometry above the feedline chip. Three different resonator lengths are chosen for the LaAlO$_3$ and TiO$_2$ substrates, respectively, and one length for the MgO substrates. All resonators have a center conductor width of $W = \SI{120}{\micro m}$ and a gap width of $S = \SI{50}{\micro m}$. 
Microwave transmission spectra are then obtained using a vector network analyzer (VNA) for frequencies up to 20~GHz while the resonator is cooled in a $^4$He cryostat with variable-temperature insert (VTI), down to a base temperature around $T \approx \SI{1.4}{K}$.\cite{Wiemann2015} 

In Fig.\ \ref{Fig:Spectra} an exemplary spectrum of a resonator on TiO$_2$ is displayed.
The resonance dips can be clearly identified. The number of possible discrete discernible resonances depends strongly on the length $l$ of the resonator and the permittivity of the studied sample, according to equation~(\ref{Eq:ResFreq}). For further analysis the resonance dips are fitted with a Lorentzian function which directly yields the resonance frequency $\nu_n$, as depicted in the insets (a), (b) in Fig.\ \ref{Fig:Spectra}.

\begin{figure}
	\centering
	\includegraphics[width=\columnwidth]{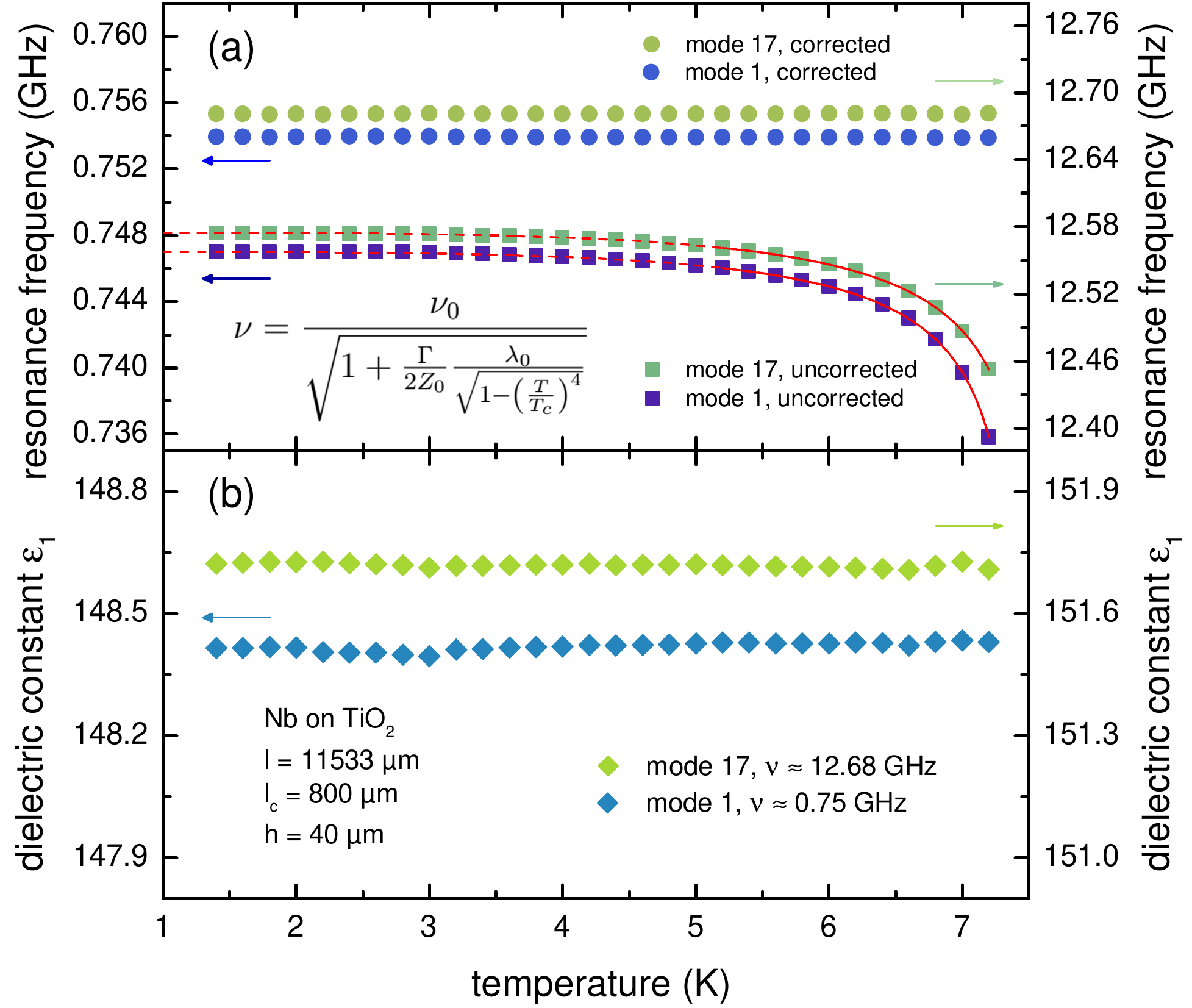}
	\caption{(a) Resonance frequencies of the fundamental ($n=1$) and $n=17$ mode for temperatures up to $T = 7.2$ K of a resonator chip on TiO$_2$. Displayed are the uncorrected resonance frequencies, where the temperature-dependent penetration depth of Nb causes a shift and the corrected resonance frequencies with the effects of the superconductor subtracted. For temperatures close to $T_c$, the displayed analytical equation is applicable and can describe the frequency shift due to the change of the impedance. After correction the permittivity can be directly calculated, as displayed in (b).}
	\label{Fig:Resonance_freq_eps_merged}
\end{figure}

These identified resonance dips are then tracked in temperature-dependent measurements up to a temperature of 9 K. In Fig.\ \ref{Fig:Resonance_freq_eps_merged}(a) the resonance frequencies of the fundamental mode as well a high harmonic of this resonator are shown. In the lower part of this panel the uncorrected frequencies can be seen. Upon increasing temperature, they first remain constant before shifting to lower frequencies. Since the permittivity of TiO$_2$ is temperature-independent within this range, this temperature dependence of the resonator has to be caused by the temperature-dependent superconducting properties of the Nb thin film. 
This can be modeled based on the London equations, which give an expression for the so-called London penetration depth.\cite{London1935} The penetration depth $\lambda$ is temperature dependent and diverges when approaching the critical temperature $T_c$. For temperatures close to $T_c$ of the Nb it can be described with an empirical equation \cite{Tinkham1996}
\begin{align}
\lambda(T) = \frac{\lambda_0}{\sqrt{1 - \left(\frac{T}{T_c} \right)^4}} \, ,
\end{align}
with $\lambda_0$ the penetration depth at zero temperature. The temperature-dependent $\lambda(T)$ affects the impedance of the coplanar resonator,\cite{Wheeler1942} and the impedance in turn is related to the resonance frequency of the resonator via the effective dielectric constant, and thus a frequency shift occurs. With increasing temperature and increasing penetration depth, the resonances shift to lower frequencies. Above $T_c$ the Nb turns to its non-superconducting metallic state with high resistivity,\cite{Kittel2005} and the resonance dips in the measured transmission spectra disappear. The temperature-dependent frequency shift applies for the fundamental and harmonic modes with index $n$ and can be described with
\begin{align}
\nu_n = \frac{\nu_{0,n}}{\sqrt{1 + \frac{\Gamma}{2Z_0}\frac{\lambda_0}{\sqrt{1 - \left(\frac{T}{T_c}\right)^4}}}} \label{eq:nu_Tc}\, ,
\end{align}
with the shifted resonance frequency $\nu$, the resonance frequency $\nu_{0}$ of the unperturbed CPW for temperature $T = 0$, a geometrical factor $\Gamma$, the characteristic impedance $Z_0$, and the penetration depth $\lambda_0$ at $T = 0$. Both, $\Gamma$ and $Z_0$ are fixed parameters and calculated with the conformal mapping theory. 

The fit to the measured data in Fig.\ \ref{Fig:Resonance_freq_eps_merged}(a) shows that this approach properly describes the temperature dependence. Therefore, we can now \lq correct\rq{} the measured data by converting them to the case of the ideal conductor, or in other words a superconductor with vanishing penetration depth.
These \lq corrected\rq{} resonance frequencies are shown in the upper part of Fig.\ \ref{Fig:Resonance_freq_eps_merged}(a). For both modes, they are basically constant for the measured temperature range. Thus they are now cleared of the temperature-dependent influence of the Nb, and they represent the intrinsic dielectric properties of the substrate material under study.

\begin{table}
	\centering
	\begin{tabular}{rcccc}
		\hline
Dielectric & $T_{c,\text{Nb}}$ (K) & $\Delta T_{c,\text{Nb}}$ (K) & $\lambda_0$ (nm) & $\Delta \lambda_0$ (nm) \\\toprule
LaAlO$_3$ & 6.91 & 0.38 & 784.26	& 169.62  \\
TiO$_2$ &  7.50	& 0.12 & 718.63 & 137.71 \\ 
MgO & 8.15 & 0.03 & 439.09	& 53.84 \\
		\hline
	\end{tabular}
		\caption{Critical temperature $T_{c,\text{Nb}}$ and penetration depth $\lambda_0$ at $T=0$ for the	Nb resonators on the respective dielectrics obtained with fits according to the equation displayed in Fig.\ \ref{Fig:Resonance_freq_eps_merged}. The values are averaged for all according fits, and $\Delta T_{c,\text{Nb}}$, $\Delta \lambda_0$ display the respective statistical errors.}
	\label{Tab:Nb}
\end{table}

The fits following equation (\ref{eq:nu_Tc}) reveal values for the parameters $T_c$ and $\lambda_0$, and thus they give information about the superconducting properties of the Nb. In table \ref{Tab:Nb} these parameters for the Nb films on the three used dielectrics are listed. They are averaged for each material, since the bulk was sputtered first and afterwards cut and structured. 
It is well known that the quality of Nb thin films is very sensitive to the growth conditions, and therefore the superconducting properties of Nb films can vary substantially. This is indeed what we find for the three different substrates, e.g.\ $T_{c,\text{Nb}}$ ranging between 6.91~K and 8.15~K, all of them well below the value of \SI{9.2}{K} for bulk Nb. Here we point out that the Nb deposition was not optimized for the different substrates, but instead the same growth recipe was applied to all of them. This was intentional, as our goal is not to study or optimize the Nb growth on a particular dielectric of interest, but rather obtain results that are independent on the quality of a particular superconducting film.

The three cases in table \ref{Tab:Nb} show that the quality of the Nb films is lowest for LaAlO$_3$, with lowest $T_{c,\text{Nb}}$ and broadest range $\Delta T_{c,\text{Nb}}$ of the superconducting transition, and highest for MgO. \lq Lower quality\rq{} of a Nb film relates to more defects in the crystal structure, which means higher scattering rate and dc resistivity for the metallic state above $T_{c,\text{Nb}}$, and following the Ferrell-Glover-Tinkham sum rule \cite{Ferrell1958,Tinkham1959} this means reduced superfluid density and longer penetration depth.
This expected trend, namely decreasing $\lambda_0$ for increasing $T_{c,\text{Nb}}$,\cite{Collignon2017} is indeed what we find for the data in table \ref{Tab:Nb}. 

After this correction for the influence of the Nb, the permittivity of the substrate can be calculated with the conformal mapping formula which gives the results shown in Fig.\ \ref{Fig:Resonance_freq_eps_merged}(b). The permittivity of TiO$_2$ is temperature-independent and the absolute values for the fundamental and highest mode are similar, around 150. Since no temperature dependence is observable, the obtained data can be averaged over the measured temperature range.

\begin{figure}
	\centering
	\includegraphics[width=\columnwidth]{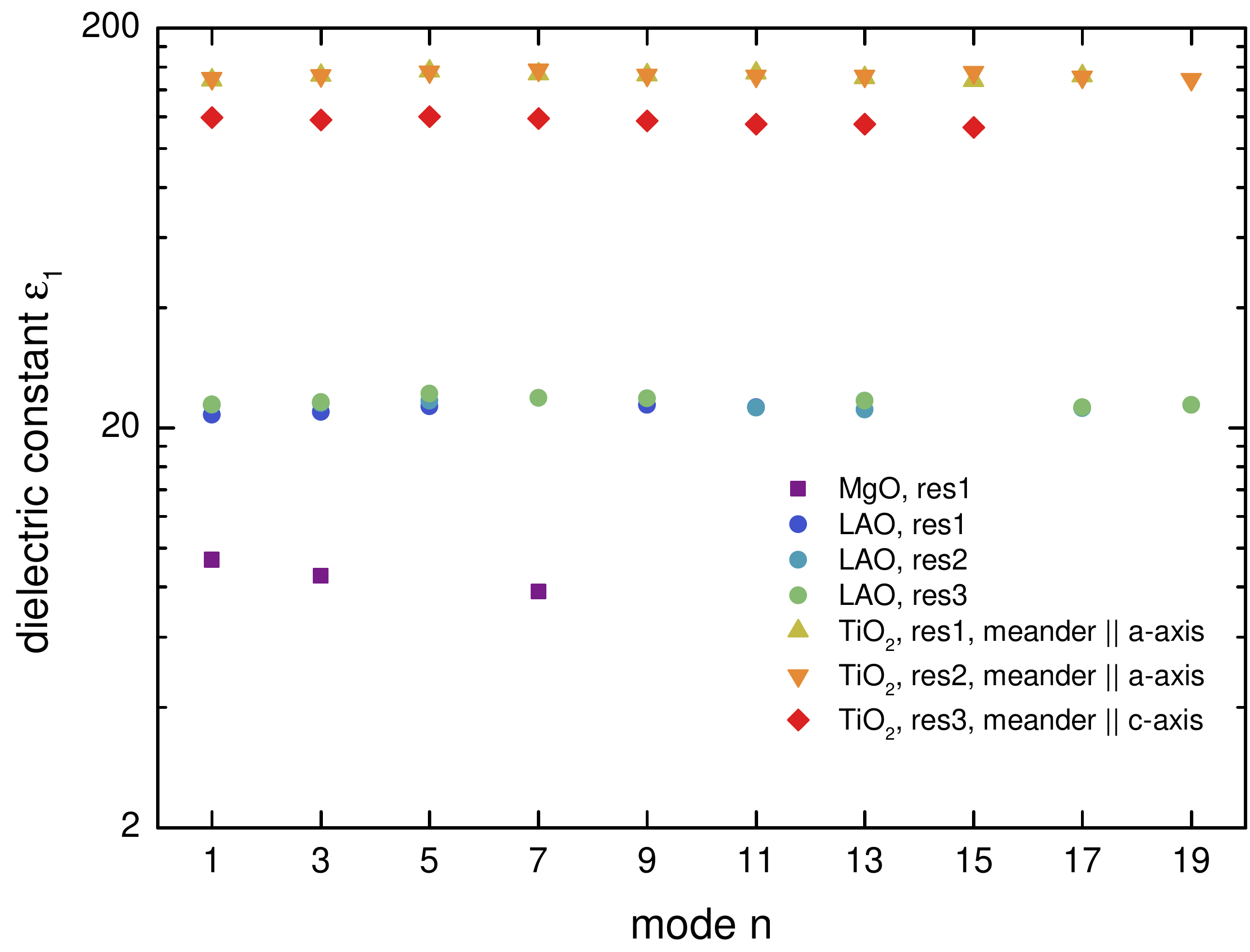}
	\caption{Dielectric constants for the  investigated samples, with each resonator operated for several harmonics with different mode $n$. The permittivities are averaged over the measured temperatures ($\SI{1.4}{K} - \SI{9}{K}$). Depending on the dielectric constant, the mode numbers correspond to frequencies up to 20~GHz.}
	\label{Fig:Eps_all}
\end{figure}

In Fig.\ \ref{Fig:Eps_all} the permittivities of all measured samples are displayed as function of the mode number. 
According to equation (\ref{Eq:ResFreq}) the mode number directly corresponds to the frequency depending on the dielectric constant of the sample, and therefore, with the upper frequency of 20~GHz in this study and the same resonator geometry for all samples, the number of accessible modes increases with increasing permittivity.

In contrast to MgO and LaAlO$_3$, the third studied material, TiO$_2$, is anisotropic. For this material, measurements were performed with the meander structure of the resonator parallel to the crystalline $a$- or $c$-axis, respectively, and consequently these experiments yield different absolute values of the permittivity. As expected, for all materials no temperature or frequency dependence is detected. For the MgO substrate experimental values for the permittivity between 8 and 10 are found, compared to the literature value of 10.\cite{Krupka1994_2} For the LaAlO$_3$ substrates they range between 21 and 24, compared to the literature value of 24.\cite{Konaka1991} 
The literature values of the permittivity for TiO$_2$ is 130 for the crystal a-axis and 250 for the c-axis.\cite{Sabisky1962,Parker1961} Here our experimental values in Fig.\ \ref{Fig:Eps_all} also show a clear difference for the two orientations of the resonator, but the resulting values of 120 and 150, differ less. This can be explained by the meander shape of the resonator, which has a large contribution of the crystal axis following the long straight sections of the meander, but the rounded parts also lead to contributions of the other crystal direction. These measurements thus show that anisotropic properties can be distinguished in principle, but for the present resonator geometry a minor contribution of the orthogonal direction is also present.

\section{Conclusions and Outlook}
This study presents a new resonant approach for characterizing dielectric bulk samples and is termed distant-flip-chip geometry. With this method multiple discrete frequencies in the microwave regime up to 20 GHz can be measured. Furthermore, the distant-flip-chip method is compatible with cryogenic temperatures. We use superconducting Nb resonators to ensure optimized measurements since Nb reduces losses at low temperatures and it needs only slight experimental preparations. Furthermore, its properties for microwave resonators are well understood. However, the Nb sputtering was not optimized for these substrate materials.\cite{Farag2019}

Nb as superconducting material limits this approach to temperatures below $\approx 7 \textrm{K}$, and this also holds for another superconductor that has been applied for related experiments and that can be easily deposited, namely Pb.\cite{Hafner2014, Koepke2014, Thiemann2018b} For several materials with unconventional properties at cryogenic temperatures, this might be sufficient, whereas in general application at higher temperatures (and higher magnetic fields) is desired. Metallic non-superconducting resonators can be operated up to room temperature and beyond,\cite{Javaheri2016} but their comparably high intrinsic losses immediately mean reduced sensitivity. Cuprate superconductors with much higher $T_c$ and critical fields thus could be an attractive alternative,\cite{Ghirri2015} but there the thin-film deposition is much more demanding.

To establish this approach, reference measurements on known dielectrics were performed. The samples are between 300 and \SI{500}{\micro m} thick and the absolute values of the permittivity are between 8 and 160. In addition, anisotropic dielectrics can be characterized, as demonstrated for the case of TiO$_2$.
This new method can now be directly transferred for measuring samples with unknown dielectric properties. Moreover, the presented geometry can be easily used for measurements at mK-temperatures and in magnetic fields.

\section*{Acknowledgments}
Financial support by Deutsche Forschungsgemeinschaft (DFG) is thankfully acknowledged.


\begin{thebibliography}{99}

\bibitem{Dimos1998} D. Dimos and C. H. Mueller,
Annu. Rev. Mater. Sci. \textbf{28} 397 (1998).

\bibitem{Brand2002}R. Brand, P. Lunkenheimer, and A. Loidl,
J. Chem. Phys. \textbf{116}, 10386 (2002).

\bibitem{Mueller1979}K. A. M\"uller and H. Burkard, 
Phys. Rev. B \textbf{19}, 3593 (1979).

\bibitem{Hess1982} H. F. Hess, K. DeConde, T. F. Rosenbaum, and G. A. Thomas,
Phys. Rev. B \textbf{25}, 5578 (1982).

\bibitem{Hering2007} M. Hering, M. Scheffler, M. Dressel, and H. v. L\"ohneysen, 
Phys. Rev. B \textbf{75}, 205203 (2007).

\bibitem{Gorshunov2016}B. P. Gorshunov, V. I. Torgashev, E. S. Zhukova, V. G. Thomas, M. A. Belyanchikov, C. Kadlec, F. Kadlec, M. Savinov, T. Ostapchuk, J. Petzelt, J. Prokleska, P. V. Tomas, E. V. Pestrjakov, D. A. Fursenko, G. S. Shakurov, A. S. Prokhorov, V. S. Gorelik, L. S. Kadyrov, V. V. Uskov, R. K. Kremer, and M. Dressel, 
Nat. Commun. \textbf{7}, 12842 (2016).

\bibitem{Krupka2006} J. Krupka,
Meas. Sci. Technol. \textbf{17}, R55 (2006).

\bibitem{Kent1994}G. Kent, 
in Proceedings of Conference on Precision Electromagnetic Measurements Digest (1994), pp. 352-353.

\bibitem{Krupka1994}J. Krupka, D. Cros, M. Aubourg, and P. Guillon, 
IEEE Trans. Microw. Theory Tech. \textbf{42}, 56 (1994).

\bibitem{Jacob2002}M. V. Jacob, J. Mazierska, K. Leong, and J. Krupka, 
IEEE Trans. Microw. Theory Tech. \textbf{50}, 474 (2002).

\bibitem{Frunzio2005}L. Frunzio, A. Wallraff D. Schuster, J. Majer, and R. Schoelkopf, 
IEEE Trans. Appl. Supercond. \textbf{15}, 860 (2005).

\bibitem{Hammer2007}G. Hammer, S. Wuensch, M. Roesch, K. Ilin, E. Crocoll, and M. Siegel, 
Supercond. Sci. Technol. \textbf{20}, S408 (2007).

\bibitem{Zmuidzinas2012} J. Zmuidzinas,
Annu. Rev. Condens. Matter Phys. \textbf{3}, 169 (2012).

\bibitem{Scheffler2013}M. Scheffler, K. Schlegel, C. Clauss, D. Hafner, C. Fella, M. Dressel, M. Jourdan, J. Sichelschmidt, C. Krellner, C. Geibel, and F. Steglich, 
Phys. Status Solidi B \textbf{250}, 439 (2013).

\bibitem{Oliver2013} W. D. Oliver and P. B. Welander,
MRS Bulletin \textbf{38}, 816 (2013).

\bibitem{Scheffler2015}M. Scheffler, M. M. Felger, M. Thiemann, D. Hafner, K. Schlegel, M. Dressel, K. S. Ilin, M. Siegel, S. Seiro, C. Geibel, and F. Steglich, 
Acta IMEKO \textbf{4}, 47 (2015).

\bibitem{Ebensperger2016}N. G. Ebensperger, M. Thiemann, M. Dressel, and M. Scheffler, 
Supercond. Sci. Technol. \textbf{29}, 115004 (2016).

\bibitem{Gruenhaupt2018}L. Gr\"unhaupt, N. Maleeva, S. T. Skacel, M. Calvo, F. Levy-Bertrand, A. V. Ustinov, H. Rotzinger, A. Monfardini, G. Catelani, and I. M. Pop,
Phys. Rev. Lett. \textbf{121}, 117001 (2018).

\bibitem{Kroll2019} J. G. Kroll, F. Borsoi, K. L. van der Enden, W. Uilhoorn, D. de Jong, M. Quintero-P\'erez, D. J. van Woerkom, A. Bruno, S. R. Plissard, D. Car, E. P. A. M. Bakkers, M. C. Cassidy, and L. P. Kouwenhoven,
Phys. Rev. Applied \textbf{11}, 064053 (2019).

\bibitem{Kubo2010} Y. Kubo, F. R. Ong, P. Bertet, D. Vion, V. Jacques, D. Zheng, A. Dr\'eau, J.-F. Roch, A. Auffeves, F. Jelezko, J. Wrachtrup, M. F. Barthe, P. Bergonzo, and D. Esteve,
Phys. Rev. Lett. \textbf{105}, 140502 (2010).

\bibitem{Driessen2012}E. F. C. Driessen, P. C. J. J. Coumou, R. R. Tromp, P. J. de Visser, and T. M. Klapwijk,
Phys. Rev. Lett. \textbf{109}, 107003 (2012).

\bibitem{Ghirri2015} A. Ghirri, C. Bonizzoni, D. Gerace, S. Sanna, A. Cassinese, and M. Affronte, Appl. Phys. Lett. 106, 184101 (2015).

\bibitem{Thiemann2018a}M. Thiemann, M. H. Beutel, M. Dressel, N. R. Lee-Hone, D. M. Broun, E. Fillis-Tsirakis, H. Boschker, J. Mannhart, and M. Scheffler, 
Phys. Rev. Lett. \textbf{120}, 237002 (2018).

\bibitem{LevyBertrand2019}F. Levy-Bertrand, T. Klein, T. Grenet, O. Dupr\'e, A. Beno\^it, A. Bideaud, O. Bourrion, M. Calvo, A. Catalano, A. Gomez, J. Goupy, L. Gr\"unhaupt, U. v. Luepke, N. Maleeva, F. Valenti, I. M. Pop, and A. Monfardini,
Phys. Rev. B \textbf{99}, 094506 (2019).

\bibitem{Megrant2012} A. Megrant, C. Neill, R. Barends, B. Chiaro, Y. Chen, L. Feigl, J. Kelly, E. Lucero, M. Mariantoni, P. J. J. O'Malley, D. Sank, A. Vainsencher, J. Wenner, T. C. White, Y. Yin, J. Zhao, C. J. Palmstr\o m, J. M. Martinis, and A. N. Cleland,
Appl. Phys. Lett. \textbf{100}, 113510 (2012).

\bibitem{OConnell2008} A. D. O'Connell, M. Ansmann, R. C. Bialczak, M. Hofheinz, N. Katz, Erik Lucero,
C. McKenney, M. Neeley, H. Wang, E. M. Weig, A. N. Cleland, and J. M. Martinis,
Appl. Phys. Lett. \textbf{92}, 112903 (2008).

\bibitem{Wisbey2019} D. S. Wisbey, M. R. Vissers, J. Gao, J. S. Kline, M. O. Sandberg, M. P. Weides, M. M. Paquette, S. Karki, J. Brewster, D. Alameri, I. Kuljanishvili, A. N. Caruso, D. P. Pappas,
J. Low Temp. Phys. \textbf{195} 474 (2019).

\bibitem{Ebensperger2019} N. G. Ebensperger, B. Ferdinand, D. Koelle, R. Kleiner, M. Dressel, and M. Scheffler,
Rev. Sci. Instrum. \textbf{90}, 114701 (2019).

\bibitem{Krupka1999}J. Krupka, K. Derzakowski, A. Abramowicz, M. E. Tobar, and R. G. Geyer, IEEE Trans. Microw. Theory Tech. \textbf{47}, 752 (1999).

\bibitem{BakerJarvis2006}J. Baker-Jarvis, M. D. Janezic, and J. Krupka, in 2006 International Conference on Microwaves, Radar Wireless Communications (2006), pp. 1093–1096.

\bibitem{BakerJarvis2010}J. Baker-Jarvis, M. D. Janezic, and D. C. Degroot, 
IEEE Instru. Meas. Mag \textbf{13}, 24 (2010).

\bibitem{Farag2019} A. Farag, M. Ubl, A. Konzelmann, M. Hentschel, and H. Giessen,
Opt. Express \textbf{27}, 25012 (2019).

\bibitem{Bothner2012a} D. Bothner, T. Gaber, M. Kemmler, D. Koelle, R. Kleiner, S. W\"unsch, and M. Siegel,
Phys. Rev. B \textbf{86}, 014517 (2012).

\bibitem{Goetz2016} J. Goetz, F. Deppe, M. Haeberlein, F. Wulschner, C. W. Zollitsch, S. Meier, M. Fischer, P. Eder, E. Xie, K. G. Fedorov, E. P. Menzel, A. Marx, and R. Gross,
J. Appl. Phys. \textbf{119}, 015304 (2016).

\bibitem{Goeppl2008}M. G\"{o}ppl, A. Fragner, M. Baur, R. Bianchetti, S. Filipp, J. M. Fink, P. J. Leek, G. Puebla, L. Steffen, and A. Wallraff,
J. Appl. Phys. \textbf{104}, 113904 (2008).

\bibitem{Bothner2012b} D. Bothner, C. Clauss, E. Koroknay, M. Kemmler, T. Gaber, M. Jetter, M. Scheffler, P. Michler, M. Dressel, D. Koelle, and R. Kleiner,
Appl. Phys. Lett. \textbf{100}, 012601 (2012).

\bibitem{Rausch2018}D. S. Rausch, M. Thiemann, M. Dressel, D. Bothner, D. Koelle, R. Kleiner, and M. Scheffler, 
J. Phys. D \textbf{51}, 465301 (2018).

\bibitem{Simons2001}R. N. Simons, 
Coplanar Waveguide Circuits, Components, and Systems (Wiley-IEEE Press, 2001).

\bibitem{Ghione1987}G. Ghione and C. U. Naldi, 
IEEE Trans. Microw. Theory Tech. \textbf{35}, 260 (1987).

\bibitem{Krupka1994_2}J. Krupka, R. G. Geyer, M. Kuhn, and J. H. Hinken, 
IEEE Trans. Microw. Theory Tech. \textbf{42}, 1886 (1994).

\bibitem{Konaka1991}T. Konaka, M. Sato, H. Asano, and S. Kubo, 
J. Supercond. \textbf{4}, 283 (1991).

\bibitem{Sabisky1962}E. S. Sabisky and H. J. Gerritsen, 
J. Appl. Phys. \textbf{33}, 1450 (1962).

\bibitem{Parker1961}R. A. Parker, 
Phys. Rev. \textbf{124}, 1719 (1961).

\bibitem{Wiemann2015} Y. Wiemann, J. Simmendinger, C. Clauss, L. Bogani, D. Bothner, D. Koelle, R. Kleiner, M. Dressel, and M. Scheffler,
Appl. Phys. Lett. \textbf{106}, 193505 (2015).

\bibitem{London1935}F. London and H. London, 
Proc. Roy. Soc. A \textbf{149}, 71 (1935).

\bibitem{Tinkham1996}M. Tinkham, 
Introduction to Superconductivity (McGraw Hill, 1996).

\bibitem{Wheeler1942}H. A. Wheeler, 
Proc. IRE \textbf{30}, 412 (1942).

\bibitem{Kittel2005}C. Kittel, 
Introduction to Solid State Physics (John Wiley \& Sons, Inc., 2005).

\bibitem{Ferrell1958}R. A. Ferrell and R. E. Glover, III,
Phys. Rev. \textbf{109}, 1398 (1958).

\bibitem{Tinkham1959}M. Tinkham and R. A. Ferrell,
Phys. Rev. Lett. \textbf{2}, 331 (1959).

\bibitem{Collignon2017}C. Collignon, B. Fauqué, A. Cavanna, U. Gennser, D. Mailly, and K. Behnia, Phys. Rev. B \textbf{96}, 224506 (2017).

\bibitem{Hafner2014} D. Hafner, M. Dressel, and M. Scheffler, 
Rev. Sci. Instrum. \textbf{85}, 014702 (2014).

\bibitem{Koepke2014} M. K\"opke and J. Weis, 
Physica C \textbf{506}, 143 (2014).

\bibitem{Thiemann2018b}M. Thiemann, M. Dressel, and M. Scheffler,
Phys. Rev. B \textbf{97}, 214516 (2018).

\bibitem{Javaheri2016} M. Javaheri Rahim, T. Lehleiter, D. Bothner, C. Krellner, D. Koelle, R. Kleiner, M. Dressel, and M. Scheffler, 
J. Phys. D: Appl. Phys. \textbf{49}, 395501 (2016).


\end{thebibliography}
\end{document}